\documentstyle[10pt,epsfig]{article}
               
\begin{document}

\begin{center}
{\Large {\bf{Measurements of the electromagnetic form factor of the Proton at JLab}}}
\vspace{.2in}

{\large {Vina Punjabi}\\ }
{\it Norfolk State University, Norfolk, VA, 23504, USA }\\
{\small \it email: punjabi@jlab.org}
\vspace{.1in}\\
{\large {C.F. Perdrisat}\\}
  {\it College of William and Mary, Williamsburg, VA 23187, USA }\\
  {\small \it email: perdrisa@jlab.org}
\vspace{.1in}\\
and 
{\large {the Jefferson Lab Hall~A and G$_{Ep}(III)$ Collaborations}}\\
  {\it 12000 Jefferson Avenue, Newport News, VA 23666, USA}

\end{center}

\begin{abstract}
\noindent
{The ratio of the proton's elastic electromagnetic form factors  
$G_{Ep}/G_{Mp}$ was obtained by 
measuring $P_{t}$ and $P_{\ell}$, the transverse and longitudinal 
recoil proton polarization, respectively. For the elastic reaction
$\vec e p \rightarrow e\vec p$, $G_{Ep}/G_{Mp}$ is proportional to 
$P_t/P_{\ell}$.  The simultaneous measurement of $P_{t}$ and $P_{\ell}$ in a polarimeter 
reduces systematic uncertainties. The results for the ratio $G_{Ep}/G_{Mp}$ measured in Hall A so far 
show a systematic decrease with increasing $Q^2$, indicating  for the 
first time a definite difference in the distribution of charge and magnetization 
in the proton. Together these experiments cover the Q$^{2}$- range of 0.5 to 5.6 GeV$^{2}$.
A new experiment is currently being prepared, to extend the Q$^2$-range to 9 
GeV$^2$ in Hall C.}
\end{abstract}

\section{Introduction}

The first direct evidence that the proton has an internal structure came from a measurement of  
its anomalous magnetic moment 70 years ago by O. Stern\cite{stern}; it is 2.79 times 
larger than that of a Dirac particle of the same mass. The first measurement of the charge 
radius of the proton, by Hofstadter\cite{hofstader} {\it{et al.}}, yielded a value of 0.8 fm  
quite close to the modern value. Although the structure of the proton has been taken for well 
known until recently, the experimental results to be reported here show that it held 
secrets which are only now being revealed. 

The nucleon elastic form factors describe the internal structure of
the nucleon; in the non-relativistic limit, for small
four-momentum transfer squared, Q$^{2}$, they are Fourier transforms of 
the charge and magnetization distributions in the nucleon. In the Breit frame 
the hadron electromagnetic 4-vector current $J_{\mu}$ has time- and space components 
proportional to the Sachs form factors G$_{Ep}$ and G$_{Mp}$, respectively. 
Hence, in this frame, it is generally true that the electric and magnetic form factors 
$G_{Ep}$ and $G_{Mp}$ are the Fourier transform of the charge and magnetization distributions, 
respectively. The difficulties associated with the calculations of the charge and magnetization 
distributions in the laboratory have been discussed recently by Kelly~\cite{kelly}.

The unpolarized elastic $ep$ cross section can be written in terms of the 
Sachs form factors $G_{Ep}$ and $G_{Mp}$ :
\begin{center}
\begin{equation}
\frac{d \sigma}{d \Omega}=\frac{{\alpha}^{2}E_{e}^{^{\prime}}\cos^2 \frac{
\theta_{e}}{2}}{4E_e^3\sin^4\frac{\theta_e}{2}} \left[G_{Ep}^2+\frac{\tau}{
\epsilon}G_{Mp}^2\right]\left(\frac{1}{1+\tau}\right),  \label{eq:mott}
\label{eq:xngegm}
\end{equation}
\end{center}
\noindent 
where $\tau=Q^{2}/4M_{p}^{2}$, $\epsilon=\left[1+2(1+\tau)\tan^2(\frac{\theta_e}{2})\right]^{-1}$
is the polarization of the virtual photon, with values between 0 and 1, 
$E_e$ and $E^{\prime}_{e}$ are the energies of the incident
and scattered electron, respectively, and $\theta_e$ is the electron scattering angle in the laboratory frame.
$G_{Ep}$ and $G_{Mp}$ can be extracted from cross section measurements made at fixed Q$^{2}$, and over a
range of $\epsilon $ values, with the Rosenbluth separation method. 

Fig. 1 shows previous results of $G_{Ep}$ and  
$G_{Mp}$ obtained by Rosenbluth separations, plotted as the ratios 
$G_{Ep}/G_D$ and $ G_{Mp}/\mu_{p} G_D$ versus $Q^2$,
up to 6~GeV$^2$. Here $G_D=(1 + Q^2/m_D^2)^{-2}$ is 
the dipole form factor, with the constant $m_D^2$ empirically determined 
to be 0.71 GeV$^2$. For $Q^2 <$ 1 GeV$^2$, the uncertainties for both 
$G_{Ep}$ and $G_{Mp}$ are only a few percent, and
one finds that $G_{Mp}/ \mu_{p} G_D \simeq G_{Ep}/G_D \simeq 1 $. 
For $G_{Ep}$ above $Q^2$ = 1 GeV$^2$, the large uncertainties 
and the scatter in results between different 
experiments, as seen in Fig.~1, illustrate the difficulties in obtaining 
$G_{Ep}$ by the Rosenbluth separation method. In contrast, the uncertainties 
for $G_{Mp}$ remain small up to $Q^2$ = 31.2 GeV$^2$\cite{sill}.
As seen in Eq.~(\ref{eq:xngegm}) the $G_{Mp}$ part of the cross section, which is about 
$\mu_p$ times larger than the $G_{Ep}$ part, is also multiplied by $\tau$; therefore, as $Q^2$ 
increases, the cross section becomes dominated by the $G_{Mp}$ term, making the extraction of $G_{Ep}$ 
more difficult by the Rosenbluth separation method.  

\begin{figure}[h]
\begin{minipage}[b]{0.45\linewidth}
\begin{center}
\epsfxsize=\textwidth
\epsfig{file=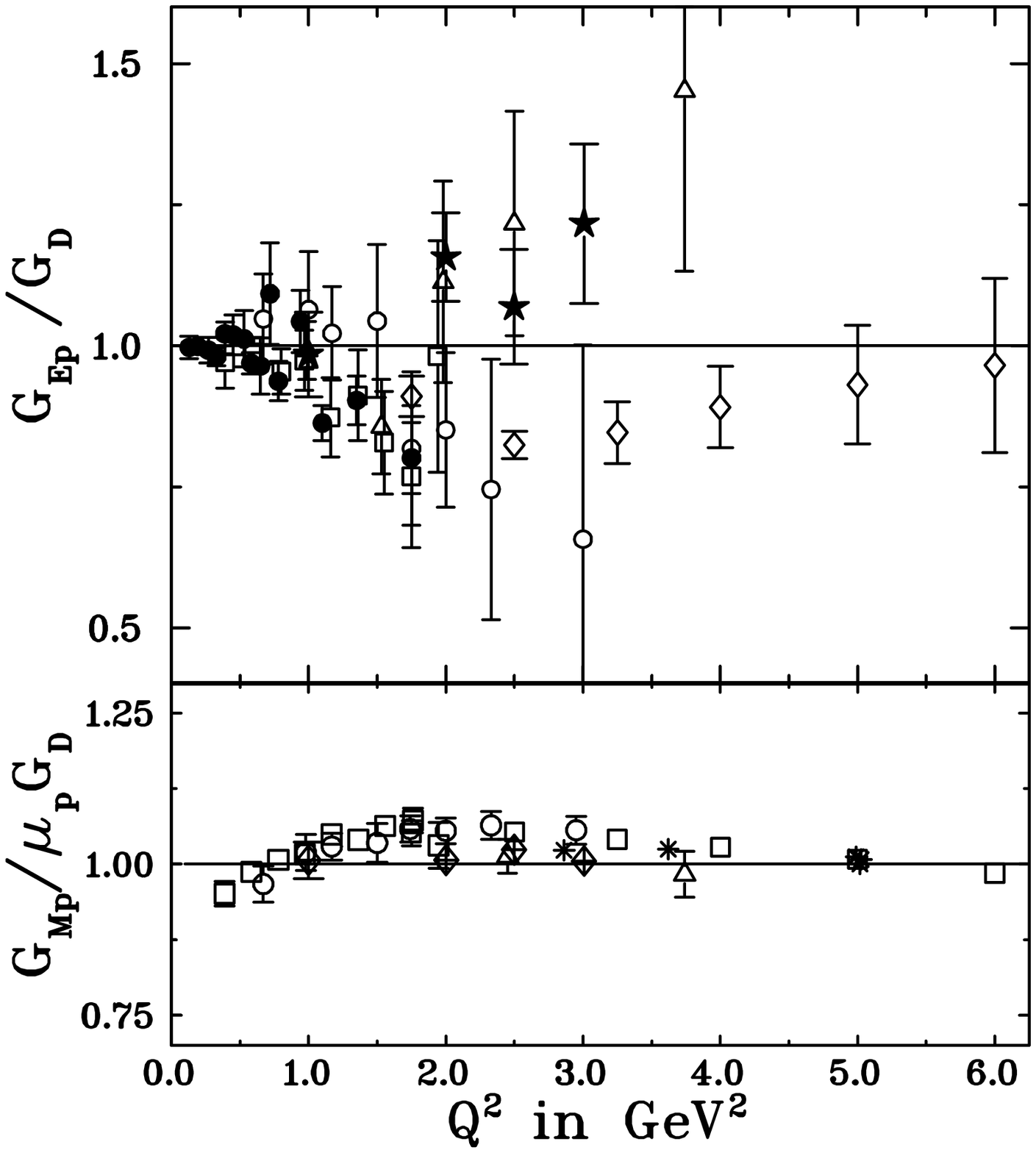,height=2.5in}
\caption[]{World data prior to 1998 for $G_{Ep}/G_D$ (top panel) and $G_{Mp}/\mu_p G_D$ (bottom panel) versus $Q^2$.
Refs. Litt {\it et al.}\cite{litt}$\triangle$, Berger {\it et al.} \cite{berger}$\Box$,  
Price {\it et al.} \cite{price}$\bullet$, Bartel {\it et al.} \cite{bartel}$\circ$, Walker 
{\it et al.} \cite{walker}$\star$, Andivahis {\it et al.} \cite{andivahis}$\Diamond$ and Sill {\it et al.} 
\cite{sill}$\ast$}
\label{fig:gepgd_gmpgd}
\end{center}
\end{minipage}\hfill
\begin{minipage}[b]{0.45\linewidth}
\begin{center}
\epsfxsize=\textwidth
\epsfig{file=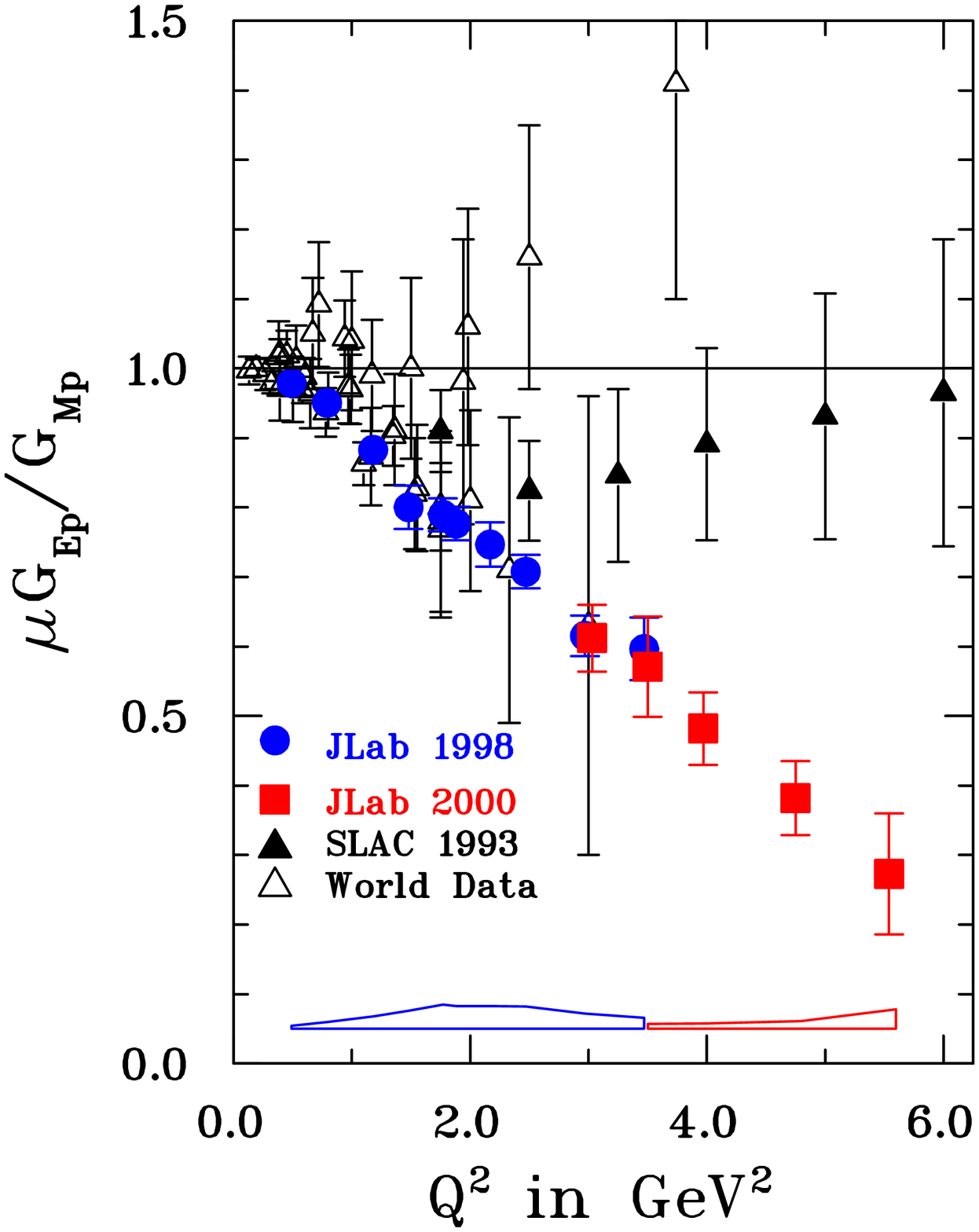,height=3.0in}
\caption{ \it  \small {The JLab data as $G_{Ep}/G_{MP}$: 93-027 (dots), 99-007 (squares);
the new data deviates strongly from the dipole form factor value of 1.}}
\label{fig:gepgd}
\end{center}
\end{minipage}\hfill
\end{figure}

The JLab results have been obtained by measuring the recoil proton
polarization in $\vec{e}p\rightarrow e\vec{p}$~\cite{Jones00,Gayou02} instead of the cross section.
For one-photon exchange, in the $\vec ep\rightarrow e\vec p$ reaction, the 
scattering of longitudinally 
polarized electrons results in a transfer of polarization to the recoil proton 
with only two non-zero components, $P_{t}$ perpendicular to, and $P_{\ell }$  
parallel to the proton momentum in the scattering plane. For 100$\%$ polarized electrons, the 
polarizations are \cite{akh1,dombey,akh2,arnold}:
\begin{center}
\begin{equation}
I_0P_t = -2\sqrt{\tau \left (1+\tau \right )}G_{E_p}G_{M_p}\tan \frac{\theta
_e}{2}  \label{eq:pol_transft}
\end{equation}
\begin{equation}
I_0P_{\ell} = \frac{1}{m_p}\left(E+E^{\prime} \right)\sqrt{\tau
\left (1+\tau \right )}G_{M_p}^2 \tan ^2\frac{\theta _e}{2}
\label{eq:pol_transfl}
\end{equation}
\end{center}
where $I_0 \propto G_{E_p}^2 + \frac{\tau}{\epsilon}G_{M_p}^2$. Measuring 
simultaneously these two components and
taking their ratio gives the ratio of the form factors:\\
\begin{center}
\vspace*{-0.3in}
\begin{equation}
\frac{G_{Ep}}{G_{Mp}}=-\frac{P_{t}}{P_{\ell}}\frac{(E+E^{\prime })} 
{2m_{p}}\tan (\frac{\theta _{e}}{2}) 
\label{eq:ratio}
\end{equation}
\end{center}
The form factor ratio $G_{Ep}/G_{Mp}$ at a given Q$^2$ can be obtained without
change of beam energy or detector angle, eliminating important sources of
systematic uncertainties; the principal source of systematic uncertainty
remaining comes from the need to account for the precession of the spin in the 
spectrometer detecting the recoil proton accurately.

\section{Experiments}

In 1998 $G_{Ep}/G_{Mp}$ was measured at JLab for Q$^{2}$ from 0.5 to 3.5~GeV
$^{2}$~\cite{Jones00}. Protons
and electrons were detected in coincidence in the two high-resolution
spectrometers (HRS) of Hall A. The polarization of the recoiling 
proton was obtained from the asymmetry of the azimuthal distribution after
rescattering the proton in a focal plane polarimeter(FPP) with graphite 
analyzer.

In  2000 new measurements were made at 
Q$^{2}$ = 4.0, 4.8 and 5.6~GeV$^{2}$ with overlap points at Q$^{2}$ = 3.0 and
3.5~GeV$^{2}$~\cite{Gayou02}. To extend the measurement to these
higher Q$^2$, two changes were made. First, to increase the 
figure-of-merit of the FPP, a CH$_{2}$ analyzer was used; the
thickness was increased from 50 cm of graphite to 100~cm of CH$_{2}$ (60~cm
for Q$^{2}$ = 3.5 GeV$^{2}$). Second, the electrons were detected in a 
lead-glass calorimeter with
9 columns and 17 rows of $15\times 15 \times 35$ cm$^{3}$ blocks 
placed so as to achieve complete solid angle matching with the 
HRS detecting the proton. At the largest Q$^{2}$ the solid
angle of the calorimeter was 6 times that of the HRS. 

The combined results from both experiments are plotted in 
Fig.~\ref{fig:gepgd} as the ratio $\mu_{p}G_{Ep}/G_{Mp}$. If the 
$\mu _{p}G_{Ep}/G_{Mp}$-ratio continues the observed linear decrease with the 
same slope, it will cross zero at Q$^{2}\approx 7.5$~GeV$^{2}$.

\section{Results and Discussion}

In exclusive electron scattering at high $Q^2$, the dominant degrees of 
freedom of the nucleon are the three valence quarks.  This is the regime 
where the perturbative QCD (pQCD) theory can be applied \cite{brodsky}.  
At moderate $Q^2$ values, the Vector Meson Dominance (VMD) 
model \cite{iachello, hohler,gari,mergell,lomon} has been successful in describing the
nucleon form factors and hadronic interactions. At 
Q$^{2}$ between 1 and 10 GeV$^{2}$, relativistic constituent quark 
models \cite{chung,aznau,dziembowski,gamiller,cardarelli,pace,Desanctis,boffi}
currently give the best understanding of the nucleon form factors, with
the strongest dynamical input. Predicting nucleon form 
factors in the intermediate $1 < Q^2 < 20$ GeV$^2$ region, where soft 
scattering processes are still dominant compared to hard scattering, is 
very difficult.

\begin{figure}[h]
\begin{minipage}[b]{0.45\linewidth}
\begin{center}
\epsfxsize=\textwidth
\epsfig{file=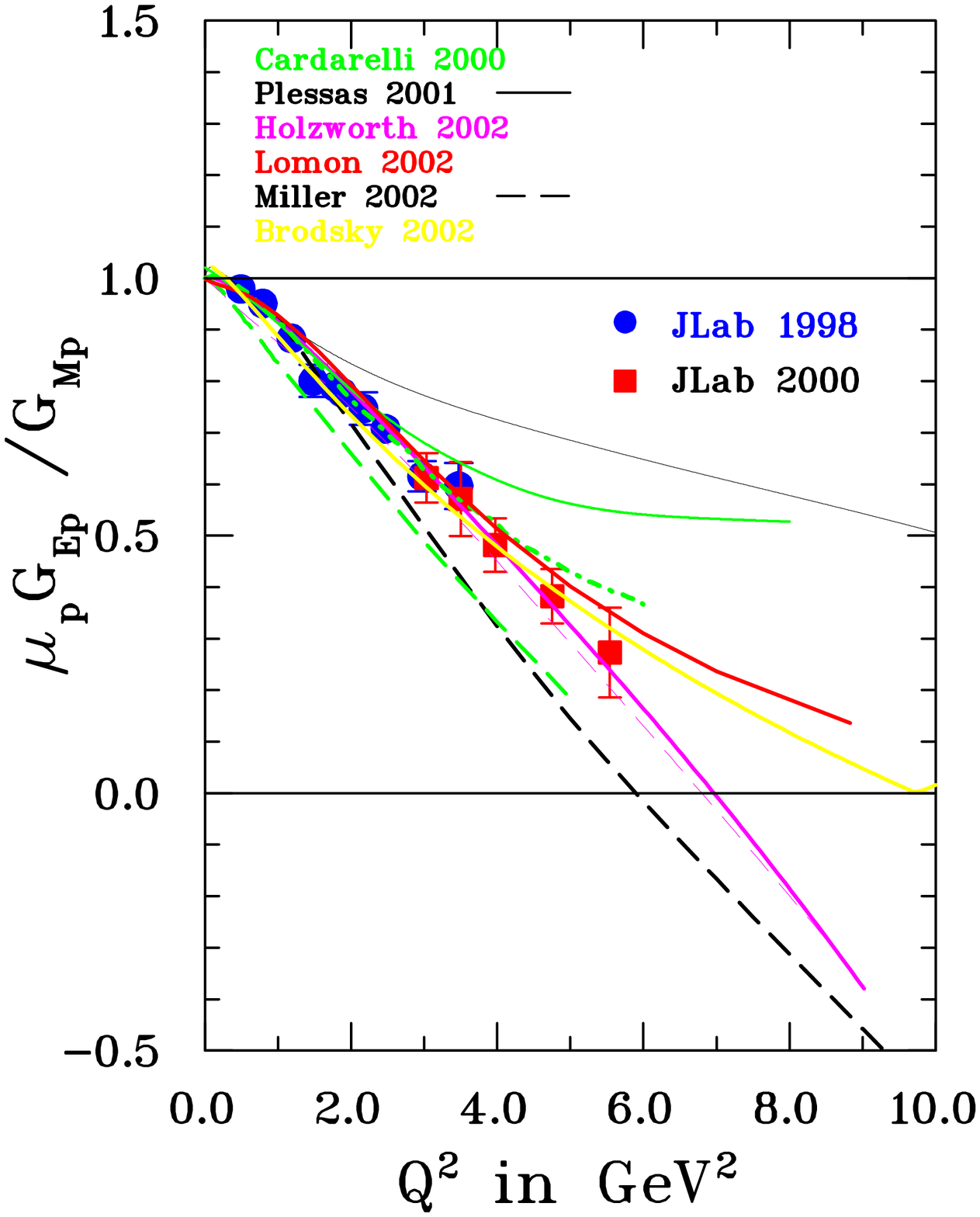,height=3.0in} 
\caption{ \it \small{Comparison of theoretical model calculations with the 
data from Ref. \cite{Jones00} (solid circles) and from \cite{Gayou02} (empty squares). 
The curves are, black thin solid \cite{boffi}, 
green dot-dashed and dashed \cite{pace}, black dashed \cite{gamiller}, red solid \cite{lomon}, 
yellow solid \cite{brodsky2002} and magenta dashed and solid \cite{holzwarth}.}}
\label{fig:gmpgd}
\end{center}
\end{minipage}\hfill
\begin{minipage}[b]{0.45\linewidth}
\begin{center}
\epsfxsize=\textwidth
\epsfig{figure=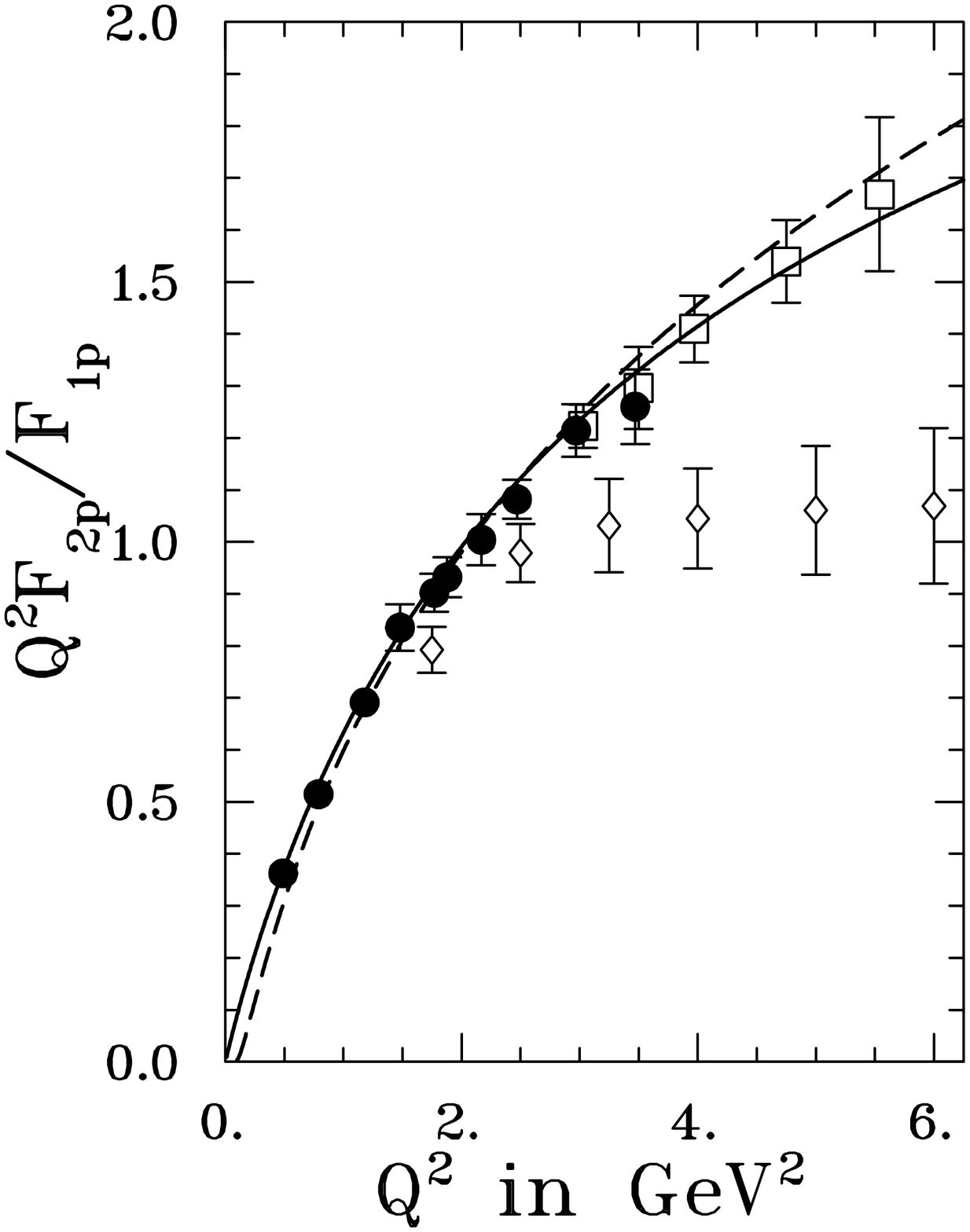, height=3.0in}
\caption{ \it \small{$Q^2 F_{2p}/F_{1p}$ versus $Q^2$. The curves shown are from Brodsky \cite{brodsky2002} 
(solid line), and from Belitsky {\it et al.} \cite{belitsky} (dashed line). The data
from Ref. \cite{Jones00} are shown as solid circles, from Ref. \cite{Gayou02} as empty squares and 
from Ref. \cite{andivahis} as 
empty diamonds.}}
\label{fig:q2f2f1}
\end{center}
\end{minipage}\hfill
\end{figure}

In Fig.~\ref{fig:gmpgd}, the JLab data are
compared with the results of various calculations that include the VMD calculation of Lomon \cite{lomon}, 
the relativistic constituent quark model (CQM) calculations \cite{gamiller,pace,boffi}, 
the soliton model calculation of Holzwarth \cite{holzwarth}, 
and a new pQCD fit from Brodsky ~\cite{brodsky2002}. In the soliton model Holzwarth\cite{holzwarth} 
applies the relativistic corrections 
due to recoil and incorporates partial coupling to vector mesons. He
uses the skyrmion as an extended object with one vector meson propagator and relativistic 
boost to the Breit frame. The result is shown in Fig.~\ref{fig:gmpgd} as the dashed and solid curves, 
corresponding to two different strength of the $\omega$-meson coupling strength, $g_{\omega}$.  
This model describes the ratio very  well over the $Q^2$ range in this experiment. 
In 2002 Lomon updated the original VMD
calculation of Gari and Kruempelman~\cite{gari} and obtained good 
agreement with the JLab data for reasonable parameters for the vector-meson masses and coupling constants.

In Fig.~\ref{fig:q2f2f1} the JLab data are shown as Q$^2$ times 
$F_{2}/F_{1}$, the ratio of the Pauli and Dirac form factors $F_2$ and $F_1$;
these are connected to the Sachs form factor as follows:
\begin{center}
\begin{equation}
F_2=\frac{G_{Mp}-G_{Ep}}{\kappa_p(1+\tau)},\hspace{0.25in}
F_1=\frac{\tau G_{Mp}+G_{Ep}}{1+\tau}
\end{equation}
\end{center}
where $\kappa_p$ is the anomalous part of the proton magnetic moment, in units of the nuclear magneton $\mu_N$.
The prediction of  pQCD is that
quenching of the spin flip form factor $F_2$ should occur at large Q$^2$, 
or equivalently helicity conservation should hold true; higher order 
contributions should make $Q^{2}F_{2}/F_{1}$ 
asymptotically constant. Unlike the SLAC~\cite{andivahis} data, the JLab data 
clearly contradict this prediction over the Q$^2$ region covered.

Recently there have been  two revisions of the pQCD prediction
for the large $Q^2$ behavior of $F_2$. In the first, Brodsky \cite{brodsky2002} argues 
that the pQCD motivated behavior of $F_2$ must contain an extra logarithmic 
term from higher twist contributions; the 3 free parameters $a$, $b$ and $c$ of the expression
$\frac{F_2p}{F_1p}=\frac{1}{1+(Q^{2}/c)\ln^b(1+Q^{2}/a)}$ were fitted in Ref. 
\cite{brodsky2002} to the data of Ref. \cite{Jones00} and \cite{Gayou02} with the result shown 
as a solid line in Fig. \ref{fig:q2f2f1}. In the second, Belitsky $\it {et~ al.}$ \cite{belitsky} 
reiterate the fact that the spin of a massless (or very light) quark 
cannot be flipped by the virtual photon of the $ep$ reaction. For a quark to 
undergo spin-flip, it must be in a state of non-zero angular momentum with projection $\mid L_z \mid = $1. 
As a result, the standard pQCD prediction for $F_{2p}$ (namely $\propto Q^{-6})$ 
becomes modified by a logarithmic term such that 
$\frac{F_{2p}}{F_{1p}}=\frac{ a}{\kappa_{p}Q^2} \ln^2(\frac{Q^2}{\Lambda^2})$, where $a$ is a 
normalization constant;    
$\Lambda$ is a cutoff constant required to suppress the infrared singularity
generated by the very soft part of the quark wave function. Although the
constant $a$ in the expression above is not determined, a fit to the data 
of this paper (augmented by the data of \cite{Gayou02}) gives $\Lambda=290$~MeV, and
$a=$0.175. The soft physics scale of the nucleon is determined by $\Lambda$; its 
size is of order of the transverse quark momentum in the nucleon. 
This fit is shown as the dashed line in Fig. \ref{fig:q2f2f1}.

\begin{figure}[h]
\begin{minipage}[b]{0.45\linewidth}
\epsfxsize=\textwidth
\epsfig{file=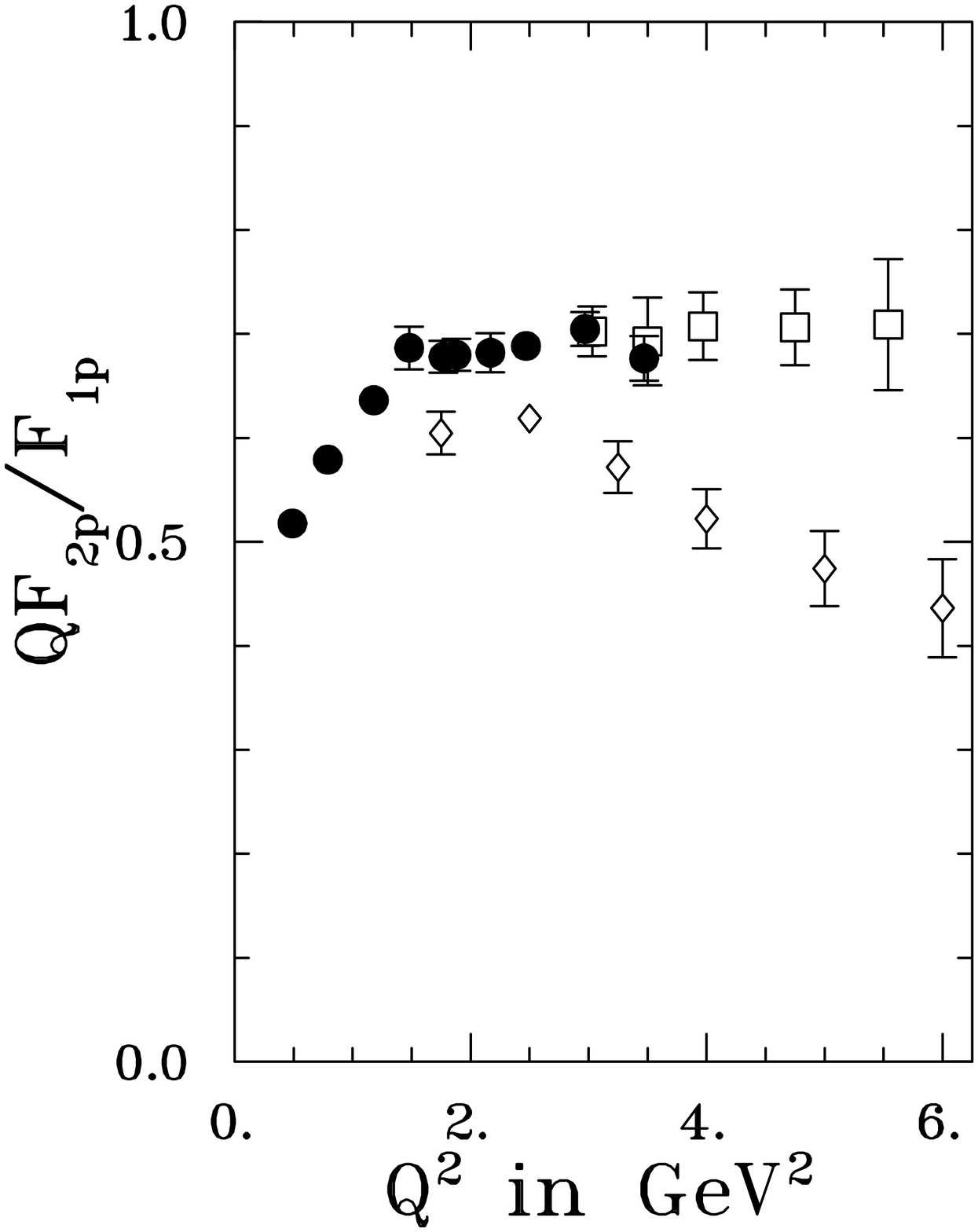, height=2.75in} 
\caption{ \it \small{The data
from Ref. \cite{Jones00} solid circles, from Ref. \cite{Gayou02} empty squares and 
from Ref. \cite{andivahis}  
empty diamonds are shown as $Q F_{2p}/F_{1p}$ vs.$Q^2$.}}
\label{fig:qf2f1}
\end{minipage}\hfill
\begin{minipage}[b]{0.45\linewidth}
\epsfxsize=\textwidth
\epsfig{figure=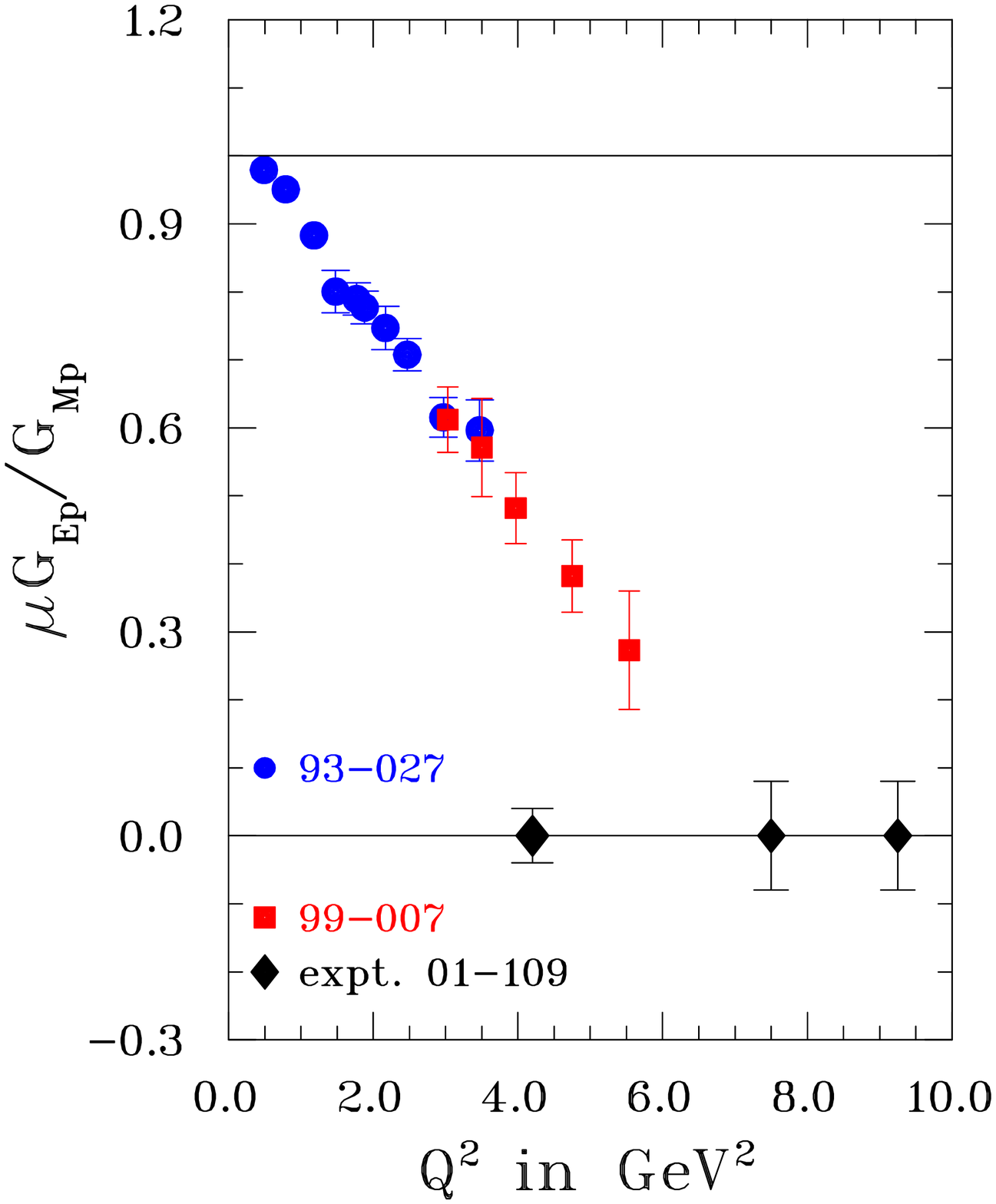, height=2.75in}
\caption{ \it \small{Predicted statistical error bars versus $Q^2$ for the 
2 new values of $Q^2$ proposed, and the control point at 4.2 $GeV^2$. 
The anticipated points are plotted arbitrarily at  $\mu G_{Ep}/G_{Mp}$=0.}}
\label{fig:gep9gev}
\end{minipage}\hfill
\end{figure}

In Fig.~\ref{fig:qf2f1} the JLab data plotted as
$QF_{2p}/F_{1p}$ show a remarkable flattening of the ratio starting at 1-2 GeV$^2$. 
Inspired by the results of this experiment,  Ralston \cite{ralston} revisited 
the calculation of the single-quark spin-flip 
amplitude responsible for the Pauli form factor in the framework of QCD. According to 
Ralston $\it {et~ al.}$ \cite{ralston,jain}, if quarks in the proton carry orbital angular momentum, 
then  $F_{2p}/F_{1p}$ should behave like $\frac{1}{\sqrt{(Q^2)}}$, 
rather than the well known pQCD prediction of ${1}\over{Q^2}$ (Ref. \cite{brodsky}). 
In a different approach, Miller and Frank ~\cite{miller} 
have shown that imposing Poincar\'{e} invariance leads to violation of the helicity 
conservation rule, which results in the behavior of $F_{2p}/F_{1p}$ observed in the JLab data.

A third phase of the $G_{Ep}/G_{Mp}$ measurements with the recoil 
polarization technique in Hall C, this time 
to 9 GeV$^2$, is planned for 2005~\cite{perdrisat}. In Fig.~\ref{fig:gep9gev} the projected error 
bars are shown at $Q^2$ of 7.5 and 9 GeV$^2$.   Also, a new Rosenbluth 
separation experiment has been done in Hall A in 2002, up to Q$^2$=4.1 GeV$^2$
~\cite{segel}; the experiment used a technique which strongly reduces 
systematic uncertainties compared to the standard Rosenbluth separation; 
the results are expected in mid-2003. A future experiment in JLab Hall A~\cite{bogdanw}, 
to measure $G_{En}$ up to 3.4 GeV$^2$, 
will significantly improve our knowledge of the nucleon form factors.

\section{Conclusion}
\label{sec:3}
The precise new JLab data on $\mu _{p}G_{Ep}/G_{Mp}$ show that this
ratio continues to drop off linearly with increasing Q$^{2}$ up to 5.6 GeV$
^{2}$. The ratio $F_2/F_1$ does not follow the $1/Q^2$ 
behavior predicted by pQCD. Thus, the Jlab data may indicate the continuing 
dominance of soft physics in the Q$^2$-range explored so far.
This behavior must be compared with the scaling of $Q^4G_{Mp}$ seen in Ref.
~\cite{sill}, starting at Q$^2\sim$ 5 GeV$^2$, which has been interpreted 
as the onset of pQCD. 
The previous discussion emphasizes the need for more and better data at higher
Q$^2$,  to challenge theoretical models in this difficult range of momentum 
transfer.

Theoretical calculations of the proton electromagnetic form factors have a 
long history. The database for three out of the four nucleon form factors is limited to 10 
GeV$^2$ or less, reaching 32 GeV$^2$ only for $G_{Mp}$. The basic physics of the 
interaction of the electromagnetic probe with the nucleon is in the difficult
region of transition between pictures of the nucleon, as a small core surrounded 
by a meson cloud, and a system of three valence quarks accompanied by gluons 
and quark-antiquark pairs described by QCD. At the lower end of this $Q^2$ range, 
the assumption that the photon interacts predominantly via an intermediate 
vector meson has been very successful; recent reconsideration of this model provides 
a quantitative description of 
the data for all four form factors. Below $Q^2$ of 10 GeV$^2$, one must use non-perturbative QCD, 
and only QCD-based phenomenological models of the nucleon are 
available. The most successful QCD based model is 
the relativistic constituent quark model, which describes the drop-off in the ratio $G_{Ep}/G_{Mp}$ observed 
in this experiment.    
At a very large, but not quantitatively defined $Q^2$, a perturbative version of 
QCD (pQCD) pioneered by Brodsky and collaborators should 
be valid. An important 
consequence of pQCD is hadron helicity conservation; in terms of the 
non-spin flip and spin flip  form factors (Dirac and Pauli), pQCD has
generally been understood to predict a faster decrease with $Q^2$ for $F_{2p}$ 
than $F_{1p}$, by a factor of 1/$Q^2$. The data presented here clearly show that
the ratio $Q^2F_{2p}/F_{1p}$ is still increasing monotonically up to 5.6 GeV$^2$. 
Recently a careful re-examination of the pQCD prediction has led to the inclusion of 
a logarithmic factor and good agreement with the behavior of $F_{2p}/F_{1p}$ reported in this paper.

\section{Acknowledgments}
We thank our colleagues M. Jones, E. Brash, L. Pentchev and  
O. Gayou for their essential roles in the completion of these experiments. 
The Southeastern 
Universities Research Association manages the Thomas Jefferson National 
Accelerator Facility under DOE contract DE-AC05-84ER40150. Our research
is supported by the the Department of Energy grant
DE-FG05-89ER40525 (VP), and U.S. National Science Foundation grant PHY 99 01182 (CFP).

\end{document}